\documentstyle[12pt]{article}

\def\be{\begin{equation}}
\def\ee{\end{equation}}

\def\ba{\begin{array}{c}}
\def\ea{\end{array}}

\begin{document}

\titlepage

 \begin{center}{\Large \bf
Conditions for complex spectra in a class of ${\cal PT}$ symmetric
potentials
  }\end{center}

\vspace{10mm}

 \begin{center}
G\'eza L\'evai

 \vspace{3mm}

Institute of Nuclear Research of the Hungarian
         Academy of Sciences,
         PO Box 51, H--4001 Debrecen, Hungary
\footnote{e-mail: levai@moon.atomki.hu}

\end{center}

\vspace{5mm}

 \begin{center}
Miloslav Znojil

 \vspace{3mm}

\'{U}stav jadern\'e fyziky AV \v{C}R, 250 68 \v{R}e\v{z}, Czech
Republic\footnote{e-mail: znojil@ujf.cas.cz}

\end{center}

\vspace{5mm}

\section*{Abstract}

We study a wide class of solvable ${\cal PT}$ symmetric potentials
in order to identify conditions under which these potentials have
regular solutions with complex energy. Besides confirming previous
findings for two potentials, most of our results are new. We
demonstrate that the occurrence of conjugate energy pairs is a
natural phenomenon for these potentials. We demonstrate that the
present method can readily be extended to further potential
classes.

\vspace{5mm}

PACS  03.65.Fd, 03.65.Ge, 03.65.Nk

\newpage

\section{Introduction}

According to a rather elementary result of quantum mechanics,
bound states in a real potential well have real eigenvalues, and the
corresponding eigenfunctions are normalizable. Recently it was
shown~\cite{bb98a} that real energy eigenvalues can  appear in
one-dimensional complex (non-Hermitian) potentials if the
Hamiltonian is ${\cal PT}$ symmetric, i.e. it is invariant under
simultaneous space (${\cal P}$) and time (${\cal T})$ reflection.
The kinetic energy term $(2m)^{-1}p^2$ is obviously invariant
under this transformation, while for the potential term
$V^*(-x)=V(x)$ is postulated.

The first ${\cal PT}$ invariant potentials have been studied using
perturbation methods~\cite{Caliceti,BG} and
numerical experiments.~\cite{DB,numerb,numer,numera} Later on, a
number of solvable models have  been identified, typically as the
complexified versions of solvable problems of quantum
mechanics. They
included the complex square well~\cite{cjt98,bb99} and all the
shape-invariant~\cite{sipot}
(plus some Natanzon-class)~\cite{natanzon}
potentials.~\cite{mz99,mzpteck,mzho,mzmorse,mzlg,glmz00}
In
these cases the existence of real energy eigenvalues has been
confirmed, however, no consistent explanation has been given for
this phenomenon. The reality of the spectra has been interpreted
in terms of
certain
analytic,~\cite{BG,Mezincescu}
Lie-algebraic~\cite{Bijan,glfcav}
and perturbative~\cite{Caliceti,tri}
arguments.

One has to note that in conventional (i.e. non-${\cal PT}$
symmetric) quantum mechanics the energy eigenvalues in complex
potentials are usually complex. The conditions for the energies Re$(E)<0$
and for the regularity of the corresponding state cease to come
together, as it can be demonstrated in a simple solvable
model~\cite{dbgljms96}.
It  turned out that also the ${\cal PT}$ invariance of the potentials
does not necessarily lead to the completely real spectrum,
and complex-energy solutions have  been constructed.
Since in this case the Hamiltonian retains ${\cal PT}$ symmetry,
while the wavefunctions do not, this phenomenon has been interpreted as
the spontaneous breakdown of ${\cal PT}$ symmetry.
Examples have been found in a
quasi-exactly solvable potential~\cite{km00} and among
shape-invariant
potentials
with a strong non-Hermiticity.~\cite{ahmed,mzcompho}

These findings obviously raised
the question whether it is possible to formulate some criteria for
the appearance of complex-energy solutions in
the generic ${\cal PT}$ symmetric
potential. No general answer has been given to this question yet,
rather the conditions varied in each case.

Motivated by the developments above, here we analyze a rather
general family of ${\cal PT}$ symmetric solvable potentials to
identify conditions under which these potentials support
{\it regular} complex-energy solutions. We focus on
potentials which, in conventional quantum mechanics, are called
shape-invariant,~\cite{sipot} and which contain the majority of
the most well-known textbook examples. These potentials can
easily be made ${\cal PT}$ symmetric by complexifying them
after setting some of their parameters to imaginary values.

In this respect a particularly interesting parameter is an
{\it imaginary} coordinate shift $x\rightarrow x+{\rm i}\epsilon$.
This transformation played a key role in converting most of
the real potentials into ${\cal PT}$ symmetric ones, because it
cancelled singularities (e.g. at the origin) which allowed the
extension of radial and periodic problems to the full $x$ axis.
In conventional quantum mechanics (real) coordinate shifts
are usually irrelevant to the problem, as they do not influence
the solutions and energy eigenvalues. In ${\cal PT}$ symmetric
quantum mechanics, however, imaginary coordinate shifts
have a special role. They can be interpreted in two ways. On the
one hand, shifting potentials to lines parallel to the $x$
axis represent a special case of the procedure by which ${\cal PT}$
symmetric potentials are defined on various contours of the
{\it complex plane}, along which the solutions are
regular~\cite{bb98a,mzmorse,mzlg}. On the other hand, however, for
these relatively simple solvable cases the potential functions can be
rewritten in such a way that the coordinate shift $\epsilon$
appears in the coupling coefficients,
and the whole problem can still be
though of as a potential defined on the $x$ axis.~\cite{glmz00}

We use a simple method based on variable transformations to derive
the potentials mentioned above.~\cite{bs62,gl89} This method was
found especially suited to the analysis of the ${\cal PT}$
symmetrized versions of some shape-invariant potentials, partly
because the imaginary coordinate shift appears in it in a natural
way.~\cite{glmz00}

\section{Derivation of the potentials by variable transformation}

Our procedure is a variable transformation $x\rightarrow z(x)$, which
takes the Schr\"odinger equation
\begin{equation}
{{\rm d}^2 \psi \over {\rm d} x^2} +(E-V(x))\psi(x)=0
\label{sch}
\end{equation}
into the second-order differential equation
\begin{equation}
{{\rm d}^2 F \over {\rm d}z^2} +Q(z) {{\rm d} F \over {\rm d}
z} +R(z) F(z)=0
\label{specf}
\end{equation}
with known solutions. Typically, $F(z)$ is a special function of
mathematical physics. Selecting $F(z)$ (and with it, the $Q(z)$
and $R(z)$ functions too), and introducing the $z(x)$ re-scaling
one can {\it define} a solvable potential,
\begin{equation}
E-V(x)
=  {z'''(x) \over 2z'(x)} -{3 \over 4} \left({z''(x) \over
z'(x)}\right)^2 +(z'(x))^2\left( R(z(x)) -{1 \over 2} {{\rm d}
Q(z) \over {\rm d}z} -{1 \over 4} Q^2(z(x))\right) \ .
\label{emv}
\end{equation}
The solutions can then be expressed in terms of $Q(z)$ and
$R(z)$ defining the special function $F(z)$, and the $z(x)$
function controlling the variable transformation:
\begin{equation}
\psi(x)\sim (z'(x))^{-\frac{1}{2}}\exp\left(\frac{1}{2}
\int^{z(x)}Q(z){\rm d}z\right) F(z(x))\ .
\label{solfq}
\end{equation}

This old method~\cite{bs62} has been applied in
a systematic search for shape-invariant potentials
in the Hermitian context in Ref.,~\cite{gl89} where
the $F(z)$ special function was chosen as an orthogonal (viz., Jacobi,
generalized Laguerre and Hermite) polynomial.
The $z(x)$ function governing the variable transformation was then
determined from the direct integration of a first-order differential
equation obtained from the condition that some term on the right-hand
side of Eq. (\ref{emv}) has to account for the constant (energy) term
on its left-hand side. This requirement leads to a first-order
differential equation for $z$ of the type
\begin{equation}
\left(\frac{{\rm d}z}{{\rm d}x}\right)^2 \phi(z)=C\ ,
\label{zdiff}
\end{equation}
where $\phi(z)$ is a function of $z$ originating from $Q(z)$ and
$R(z)$. Its general solution is given by
\begin{equation}
\int \phi^{1/2}(z) {\rm d}z=C^{1/2}x+\delta \ ,
\label{zx}
\end{equation}
where, in contrast to Hermitian problems, the constant of
integration plays a special role for ${\cal PT}$ symmetric systems
as the {\it imaginary} coordinate shift.~\cite{glmz00} However,
similarly to the Hermitian case, this coordinate shift has no
effect on the energy eigenvalues.
A further important implication of this method is that
the transformation properties of the $z(x)$ function under the
${\cal PT}$ operation can easily be established, which facilitates
the enforcement of ${\cal PT}$ symmetry on the potential $V(x)$ in
Eq. (\ref{emv}). In this way whole classes of potentials can be
treated on an equal footing.

Here we specify the method for the Jacobi and the generalized
Laguerre polynomials by choosing
$Q(z)=(\beta-\alpha)(1-z^2)^{-1}-(\alpha+\beta+2)z(1-z^2)^{-1}$,
$R(z)=n(n+\alpha+\beta+1)(1-z^2)^{-1}$ and
$Q(z)=-1+(\alpha+1)/{z}$, $R(z)=n/z$, respectively.
In what follows we occasionally make reference to the a previous
study in which conditions have been derived for having {\it real}
spectra of the same ${\cal PT}$ symmetric potentials.~\cite{glmz00}.

\subsection{The PI potential family}

The generic form of the shape-invariant PI type potentials (which
are known as type A potentials elsewhere~\cite{ih51,gl89}) and their
energy spectrum is
\begin{eqnarray}
E-V(x)&=&C\left(n+\frac{\alpha+\beta+1}{2}\right)^2
+\frac{C}{1-z^2(x)}
\left[\frac{1}{4}-\left(\frac{\alpha+\beta}{2}\right)^2
-\left(\frac{\alpha-\beta}{2}\right)^2\right]
\nonumber\\
&& -\frac{2Cz(x)}{1-z^2(x)}
\left(\frac{\alpha+\beta}{2}\right)
\left(\frac{\alpha-\beta}{2}\right)\ .
\label{emvjacpi}
\end{eqnarray}
This is obtained from ({\ref{emv}) by substituting the
appropriate form of $Q(z)$ and $R(z)$ and setting
$\phi(z)=(1-z^2)^{-1/2}$ in ({\ref{zdiff}). As discussed
in,~\cite{glmz00} the four different solutions $z(x)$ of (\ref{zx})
are obtained as hyperbolic and trigonometric functions and
are either invariant under the ${\cal PT}$ operation, or change
their sign. The six PI type ${\cal PT}$ symmetric potentials
and the energy eigenvalues are listed in the first six lines of
Table 1, along with the conditions under which the eigenvalues
are complex and the corresponding solutions are regular. These
potentials are obtained from (\ref{emvjacpi}) by setting $z(x)$
and $C$ to ${\rm i}\sinh(ax+{\rm i}\epsilon)$,
$\cosh(ax+{\rm i}\epsilon)$,
$\cosh(2ax+{\rm i}\epsilon)$,
$\cos(ax+{\rm i}\epsilon)$,
$\cos(2ax+{\rm i}\epsilon)$,
$\sin(ax+{\rm i}\epsilon)$ and
$C=-a^2$, $-a^2$, $-4a^2$, $a^2$, $4a^2$, $a^2$.

Let us discuss the details with the example of the potential
appearing in the first line,
\begin{equation}
V(x)=-\left(\frac{\alpha^2+\beta^2}{2}-\frac{1}{4}\right)
\frac{a^2}{\cosh^2(ax+{\rm i}\epsilon)}
-{\rm i}a^2
\left(\frac{\alpha^2-\beta^2}{2}\right)
\frac{\sinh(ax+{\rm i}\epsilon)}{\cosh^2(ax+{\rm i}\epsilon)}\ .
\label{pisinh}
\end{equation}
(It Table 1 we used $a=1$ everywhere for simplicity.)
This potential
remains non-singular in the limit $\epsilon \to 0$.
Specifying the generic form of the solution ({\ref{solfq})
for this case we find that it is regular for $x\rightarrow\pm\infty$
(i.e. $\vert z\vert\rightarrow \infty$) if
$n<-\frac{1}{2}[{\rm Re}(\alpha+\beta)+1]$ holds, while
regularity for $x\rightarrow 0$ ($z\rightarrow \pm 1$)
requires $\epsilon\ne \frac{\pi}{2}\pm k\pi$. Combining these with
the conditions $\alpha^*=\pm\alpha$, $\beta^*=\pm\beta$ securing
${\cal PT}$ symmetry of the potential~\cite{glmz00} and the
requirement of having complex energy eigenvalues we find that
to satisfy all these criteria, either $\alpha$ or $\beta$
has to be imaginary, and also we need
$\epsilon\ne \frac{\pi}{2}\pm k\pi$, as displayed in Table 1.

This potential has also been analyzed in~\cite{ahmed} for
the special case of $\epsilon=0$.
There it is underlined that complex
energies appear if $\vert V_2\vert > V_1+\frac{1}{4}$ holds, where
$V_1$ and $V_2$ are the coefficients of the real and the imaginary
parts of the potential. This agrees with our results
(for $\epsilon=0$), according to which either $\alpha$ or $\beta$
has to be imaginary, since
$-\vert V_2\vert + V_1+\frac{1}{4}$
equals to the square of the imaginary parameter (i.e. $\alpha^2$ or
$\beta^2$) and is always negative. If both $\alpha$
and $\beta$ are imaginary, then the solutions become irregular
for $\vert z\vert\rightarrow\infty$, while if both of them are
real, then the energy eigenvalues  become real. For $\epsilon\ne0$
the two potential terms both become complex, so the decomposition of
the potential into real and imaginary component is less trivial.
We only note that for $\epsilon=\frac{\pi}{2}\pm k\pi$ the
potential becomes the real version of the singular generalized
P\"oschl--Teller potential, corresponding to the case denoted with
PI($\cosh(ax+{\rm i}\epsilon)$), setting $\epsilon=0$.

The conditions are rather similar to the other type PI
potentials, with the exception that for the trigonometric
potentials the regularity requirement of the solutions
$\psi_n(x)$ at $\vert z\vert\rightarrow \infty$ is irrelevant,
and therefore the principal quantum number $n$ can take any
number, i.e.  these potentials have infinite bound states. It
has to be noted that further analysis of the boundary conditions
is necessary if these trigonometric potentials are {\it defined}
to be periodic by extending them to the full $x$
axis.~\cite{numer,mzparadox}

\subsection{The PII potential family}

The PII (or type E~\cite{ih51,gl89}) potentials are obtained
in a similar way, taking
$\phi(z)=(1-z^2)^{-1}$ in ({\ref{zdiff}).
The four ${\cal PT}$ symmetric potentials belonging to this
class are again characterized by $z(x)$ functions of the
hyperbolic or trigonometric type, which are either invariant
under the ${\cal PT}$ operation or change sign. The generic
form of the potential and the energy spectrum for this class is
\begin{eqnarray}
E-V(x)&=&
-C\left[\left(\frac{\alpha+\beta}{2}\right)^2
+\left(\frac{\alpha-\beta}{2}\right)^2\right]
-2C\left(\frac{\alpha+\beta}{2}\right)
\left(\frac{\alpha-\beta}{2}\right)z(x)
\nonumber \\
&& +C\left(n+\frac{\alpha+\beta}{2}\right)
\left(n+\frac{\alpha+\beta}{2}+1\right)(1-z^2(x))
\ .
\label{emvjacpiiab}
\end{eqnarray}
With a parameter transformation
$s=n+(\alpha+\beta)/2$
and $\Lambda=\frac{\alpha-\beta}{2}\frac{\alpha+\beta}{2}$
the dependence on $n$ can be shifted to the constant (energy) term,
leading to the potential function
\begin{equation}
V(x)=-C s(s+1)(1-z^2(x))-2C\Lambda z(x)\ , \label{vjacpii}
\end{equation}
and the
$E=-C[(s-n)^2+\Lambda^2(s-n)^{-2}]$
energy formula.
The four ${\cal PT}$ symmetric PII type potentials are again displayed
in lines seven to ten in Table 1. These are obtained from
(\ref{vjacpii}) by choosing $z(x)=\tanh(ax+{\rm i}\epsilon)$,
$\coth(ax+{\rm i}\epsilon)$,
$-{\rm i}\cot(ax+{\rm i}\epsilon)$,
${\rm i}\tan(ax+{\rm i}\epsilon)$;
$C=a^2$, $a^2$, $-a^2$, $-a^2$; and
$\Lambda={\rm i}\lambda$, ${\rm i}\lambda$, $-\lambda$ and $\lambda$.

As an example, let us analyse in more detail the
hyperbolic Rosen--Morse potential obtained for
$z(x)=\tanh(ax+{\rm i}\epsilon)$,
\begin{equation}
V(x)=-a^2\frac{s(s+1)}{\cosh^2(ax+{\rm i}\epsilon)}
-2{\rm i}\lambda a^2\tanh(ax+{\rm i}\epsilon)\ .
\label{piitanh}
\end{equation}
(We used $a=1$ in Table 1 for these potentials too, for simplicity.)
This potential is ${\cal PT}$ invariant, if $\lambda$ is real, and
$s$ is either real or it is complex, with
Re$(s)=-1/2$.~\cite{glmz00}
From Table 1 we then read that complex energies can appear only in
this latter case, i.e. if $s=-\frac{1}{2}+{\rm i}\sigma$.
However, this contradicts the regularity requirement for the solutions at
$z\rightarrow\pm 1$, which allows regular solutions only for
real values of $s$, therefore we conclude that this potential
cannot support complex-energy solutions. This is not surprizing,
at least for $\epsilon=0$, since then $s(s+1)=-\sigma^2-\frac{1}{4}$,
i.e. the first (real) term of the potential is repulsive. For
$\epsilon\ne 0$ the situation is less clear, in a way similar to the PI
type potentials.

With the exception of the above
two examples, the potentials in Table 1
have singularities, and they can be regularized only with the use
of the non-vanishing imaginary coordinate shift i$\epsilon$.
Otherwise,
the detailed discussion is rather similar for these
potentials, except that the requirements are again more relaxed
for the trigonometric potentials, so complex-energy solutions are
possible in those cases.

\subsection{The LI potential family}

Potentials related to the generalized Laguerre polynomials are
obtained by setting
$Q(z)=-1+(\alpha+1)/{z}$ and $R(z)=n/z$.
Setting $\phi(z)=z^{-1/2}$ in ({\ref{zdiff}), the type LI potential,
i.e. the harmonic oscillator is obtained (which is referred to
as the type C family elsewhere.~\cite{ih51,gl89}) The ${\cal PT}$
symmetric version of this problem is obtained using the complex
$z(x)=\frac{C}{4}(x+{\rm i}\epsilon)^2$ function.
(We note that due to the particular form of the solutions (\ref{solfq}),
the imaginary coordinate shifts do not lead to normalizable states
for the other two shape-invariant potentials related to the
generalized Laguerre polynomials, the LII (or type
F~\cite{ih51,gl89}) Coulomb and the LII (or type B~\cite{ih51,gl89})
Morse potentials, so these have to be defined on more sophisticated
contours of the complex plane.~\cite{mzmorse,mzlg})

The ${\cal PT}$ symmetric harmonic oscillator is then obtained
from the current version of (\ref{emv}) as
\begin{equation}
E-V(x)=C\left(n+\frac{\alpha+1}{2}\right)-\frac{C}{4}z(x)
-\frac{C}{4z(x)}\left(\alpha^2-\frac{1}{2}\right)\ ,
\label{emvli}
\end{equation}
if $\alpha$ is real or imaginary.~\cite{glmz00}
Using the notation $C=2\omega$ one gets the familiar form of the
potential displayed in the last line of Table 1.
Its solutions are necessarily regular asymptotically (for
$\vert z\vert\rightarrow\infty$) and also at $x\rightarrow 0$,
if $\epsilon\ne 0$ holds. Complex-energy
solutions emerge if $\alpha$ is imaginary, provided that the
singularity of the potential is cancelled by taking $\epsilon\ne 0$.
This induces a finite and attractive centrifugal term with
complex angular momentum. This again agrees with the findings of
Refs.~\cite{mzho,mzcompho}

\section{Conclusions}

In conclusion, we have reviewed a wide class of ${\cal PT}$ symmetric
potentials and specified conditions under which they can support
complex-energy solutions. For this we had to combine various
requirements for the potential parameters. In some cases
 these requirements contradict each other, so no such potentials
can exist. Our results agree with those of Ref.~\cite{mzcompho} for the
harmonic oscillator, and  contain the findings of Ref.~\cite{ahmed}
as a special case. All our other results are new.

A general feature of the present potential class is that the functional
form of the potentials depends on the {\it squares} of the potential
parameters which can take on imaginary values
(i.e. $\alpha$, $\beta$, i$\sigma$), therefore the potentials
are insensitive to the sign of this parameter. However, this sign
appears explicitly in the energy formulae as the sign of the imaginary
component of the energy, thus the occurrence of complex conjugate energy
pairs is a necessity. From the structure of the energy formulae it is
apparent that depending on the potential parameters,
the energy eigenvalues of these potentials either all real or complex,
so they practically do not occur together at the same time.
It can be noted, however, that it might be possible to make a
single energy eigenvalue real by ``fine tuning'' the potential
parameters such that the principal quantum number $n$ is cancelled by
the real part of the potential parameters and the numerical constants.

Further systematic studies of the same potentials can  be
planned to find real spectra in complex potentials that do not
fulfill ${\cal PT}$ symmetry. Such potentials have been
sought,~\cite{cjt98,nonpt} and the present method seems rather appropriate
for such purposes. As a further possibility, a similar study of
${\cal PT}$ symmetric potential defined on bent contours of the
complex $x$ plane  represents a challenging task.

We  note that the present formalism is applicable to potentials
beyond the shape-invariant class. In particular, some ``implicit''
potentials belonging to the Natanzon class~\cite{natanzon} can
 be made ${\cal PT}$ symmetric,~\cite{glmz00} and the conditions
for real- or complex-energy regular solutions can be derived.
As an example we mention the potential in Ref.~\cite{dutt}, which
had been introduced as conditionally exactly
solvable. Later it was shown that it is just an element of the
exactly solvable Natanzon potential class.~\cite{rrprmzgl01}
The energy eigenvalues of this
potential are obtained from a cubic algebraic equation. The
complexified ${\cal PT}$ symmetric version of this potential
has been studied,~\cite{compdutt} and in the present context the
interpretation of the complex roots of the cubic equation seems
a rather intriguing question.

\section*{Acknowledgments}

This work was supported by the OTKA grant No.
T031945 (Hungary) and grant No. 1048004 of GA AS CR (Czech
Republic).

\begin{table}[htbp]
\caption{Solvable potentials with spontaneous ${\cal PT}-$symmetry
breaking.}
\begin{tabular}{ccl}\\
\hline
$V(x)$ & $E_n$ & Complex-energy \\
&& regular solutions \\
\hline
$-\frac{2(\alpha^2+\beta^2)-1}{4\cosh^2(x+{\rm i}\epsilon)}
-{\rm i}\frac{(\alpha^2-\beta^2)\sinh(x+{\rm i}\epsilon)
 }{2\cosh^2(x+{\rm i}\epsilon)}$ &
$-\left(n+\frac{\alpha+\beta+1}{2}\right)^2$ &
$\alpha$ or $\beta$ imaginary, \\
 && \ $\epsilon\ne\frac{\pi}{2}\pm k\pi$\\
$\frac{2(\alpha^2+\beta^2)-1}{4\sinh^2(x+{\rm i}\epsilon)}
+\frac{(\alpha^2-\beta^2)\cosh(x+{\rm i}\epsilon)
 }{2\sinh^2(x+{\rm i}\epsilon)}$ &
$-\left(n+\frac{\alpha+\beta+1}{2}\right)^2$ &
$\alpha$ or $\beta$ imaginary, \\
 && \ $\epsilon\ne k\pi$\\
$-\frac{4\beta^2-1}{4\cosh^2(x+\frac{{\rm i}}{2}\epsilon)}
+\frac{4\alpha^2-1}{4\sinh^2(x+\frac{{\rm i}}{2}\epsilon)}$ &
$-(2n+\alpha+\beta+1)^2$ &
$\alpha$ or $\beta$ imaginary, \\
 && \ $\epsilon\ne k\pi$\\
$\frac{2(\alpha^2+\beta^2)-1}{4\sin^2(x+{\rm i}\epsilon)}
+\frac{(\alpha^2-\beta^2)\cos(x+{\rm i}\epsilon)
 }{2\sin^2(x+{\rm i}\epsilon)}$ &
$\left(n+\frac{\alpha+\beta+1}{2}\right)^2 $&
$\alpha$ and/or $\beta$ imag., \\
 && Im$(\alpha+\beta)\ne 0$, $\epsilon\ne 0$\\
$\frac{4\beta^2-1}{4\cos^2(x+\frac{{\rm i}}{2}\epsilon)}
+\frac{4\alpha^2-1}{4\sin^2(x+\frac{{\rm i}}{2}\epsilon)}$ &
$(2n+\alpha+\beta+1)^2$ &
$\alpha$ and/or $\beta$ imag., \\
 && Im$(\alpha+\beta)\ne 0$, $\epsilon\ne 0$\\
$\frac{2(\alpha^2+\beta^2)-1}{4\cos^2(x+{\rm i}\epsilon)}
+\frac{(\alpha^2-\beta^2)\sin(x+{\rm i}\epsilon)}{
 2\cos^2(x+{\rm i}\epsilon)}$ &
$\left(n+\frac{\alpha+\beta+1}{2}\right)^2$ &
$\beta=-\alpha^*$ imag., \\
 && \ $\epsilon\ne 0$ \\
$-\frac{s(s+1)}{\cosh^2(x+{\rm i}\epsilon)}
-2{\rm i}\lambda \tanh(x+{\rm i}\epsilon)$ &
$-(s-n)^2+\frac{\lambda^2}{(s-n)^2}$ &
no such solutions \\
 &&\\
$\frac{s(s+1)}{\sinh^2(x+{\rm i}\epsilon)}
-2{\rm i}\lambda \coth(x+{\rm i}\epsilon)$ &
$-(s-n)^2+\frac{\lambda^2}{(s-n)^2}$ &
no such solutions \\
 &&\\
$\frac{s(s+1)}{\sin^2(x+{\rm i}\epsilon)}
-2{\rm i}\lambda \cot(x+{\rm i}\epsilon)$ &
$(s-n)^2+\frac{\lambda^2}{(s-n)^2}$ &
$s=-\frac{1}{2}+{\rm i}\sigma$, \\
 && \ $\epsilon\ne 0$ \\
 &&\\
$\frac{s(s+1)}{\cos^2(x+{\rm i}\epsilon)}
+2{\rm i}\lambda \tan(x+{\rm i}\epsilon)$ &
$(s-n)^2+\frac{\lambda^2}{(s-n)^2}$ &
$s=-\frac{1}{2}+{\rm i}\sigma$, \\
 && \ $\epsilon\ne 0$ \\
$\frac{\omega^2}{4}(x+{\rm i}\epsilon)^2 + (\alpha^2-\frac{1}{4})
\frac{1}{(x+{\rm i}\epsilon)^2}$ &
$2\omega(n+\frac{\alpha+1}{2})$ &
$\alpha$ imaginary, \\
 && \ $\epsilon\ne 0$ \\
\hline\\

\end{tabular}

\end{table}

\end{document}